\documentclass[aps,preprint,superscriptaddress,12pt]{revtex4}

\usepackage{amsfonts}
\usepackage{amsmath}
\usepackage{amssymb}
\usepackage{graphicx}
\usepackage{bm}
\usepackage[usenames,dvipsnames]{xcolor}

\usepackage{hyperref}
\usepackage{setspace}
\usepackage{booktabs}
\usepackage{tabularx}

\begin{document}
\noindent Note: This paper has been published in \emph{Nature Physics} \textbf{11}, 779-786 (2015). Main paper and SI: http://www.nature.com/nphys/journal/v11/n9/full/nphys3422.html\\

\title{Spectrum of Controlling and Observing Complex Networks}

\author{Gang Yan}\thanks{These authors contributed equally to this work.}
\author{Georgios Tsekenis}\thanks{These authors contributed equally to this work.}
\affiliation{Center for Complex Network Research and Department of Physics, Northeastern University, Boston, Massachusetts 02115, USA}

\author{Baruch Barzel}
\affiliation{Department of Mathematics, Bar-Ilan University, Ramat-Gan 52900, Israel}

\author{Jean-Jacques Slotine}
\affiliation{Department of Mechanical Engineering and Department of Brain and Cognitive Sciences, Massachusetts Institute of Technology, Cambridge, Massachusetts 02139, USA}

\author{Yang-Yu Liu}
\affiliation{Channing Division of Network Medicine, Brigham and Women's Hospital, Harvard Medical School, Boston, Massachusetts 02115, USA}
\affiliation{Center for Cancer Systems Biology, Dana Farber Cancer Institute, Boston, Massachusetts 02115, USA}

\author{Albert-L\'aszl\'o Barab\'asi}\thanks{Corresponding author. Email: alb@neu.edu}
\affiliation{Center for Complex Network Research and Department of Physics, Northeastern University, Boston, Massachusetts 02115, USA}
\affiliation{Center for Cancer Systems Biology, Dana Farber Cancer Institute, Boston, Massachusetts 02115, USA}
\affiliation{Department of Medicine, Brigham and Women's Hospital, Harvard Medical School, Boston, Massachusetts 02115, USA}
\affiliation{Center for Network Science, Central European University, H-1051 Budapest, Hungary}


\maketitle

\noindent\textbf{Observing and controlling complex networks are of paramount interest for understanding complex physical, biological and technological systems. Recent studies have made important advances in identifying sensor or driver nodes, through which we can observe or control a complex system. Yet, the observational uncertainty induced by measurement noise and the energy required for control continue to be significant challenges in practical applications. Here we show that the variability of control energy and observational uncertainty for different directions of the state space depend strongly on the number of driver nodes. In particular, we find that if all nodes are directly driven, control is energetically feasible, as the maximum energy increases sublinearly with the system size. If, however, we aim to control a system through a single node, control in some directions is energetically prohibitive, increasing exponentially with the system size. For the cases in between, the maximum energy decays exponentially when the number of driver nodes increases. We validate our findings in several model and real networks, arriving to a series of fundamental laws to describe the control energy that together deepen our understanding of complex systems.}

\vspace{1.5cm}
\noindent Many natural and man-made systems can be represented as networks\cite{Albert-RMP-02,Cohen-Book-10,Newman-Book-10},
where nodes are the system's components and links describe the interactions between them. Thanks to these interactions, perturbations of one node can alter the states of the other nodes\cite{Boccaletti-PR-2006,barrat2008dynamical,barzel2013universality}. This property has been exploited to control a network, i.e. to move it from an initial state to a desired final state\cite{Rugh-Book-1996,Sontag-Book-1998,Slotine-Book-91}
by manipulating the state variables of only a subset of its nodes\cite{Liu-Nature-11}. Such control processes\cite{Liu-Nature-11,Sorrentino-PRE-07,Yu-Automatica-09,Rajapakse-PNAS-11,Nepusz-NP-12,Yan-PRL-12,Sun-PRL-2013,Pasqualetti-IEEECNS-2014,Tang-PO-2012,Jia-NC-2013,Yuan-NC-2013,Ruths-Science-2014,Giulia-PRL-2014,summers2014submodularity,Tzoumas-arXiv-2015,Cornelius-NC-2013,Whalen-PRX-2015} play an important role in
the regulation of protein expression\cite{Menolascina-PlosCB-2014},
the coordination of moving robots\cite{Rahmani-SIAM-09},
and the inhibition of undesirable social contagions\cite{Acemoglu-GEB-2010}.
At the same time the interdependence between nodes means that the states of a small number of sensor nodes contain sufficient information about the rest of the network, so that we can reconstruct the system's full internal state by accessing only a few outputs\cite{Liu-PNAS-13}.
This can be utilized for biomarker design in cellular networks, or to monitor in real time the state and functionality of infrastructural\cite{Yang-PRL-2012}
and social-ecological\cite{Pinto-PRL-2012}
systems for early warning of failures or disasters\cite{Scheffer-Science-2012}.

While recent advances in driver and sensor node identification constitute unavoidable steps towards controlling and observing real
networks, in practice we continue to face significant challenges: the control of a large network may require a vast amount of energy\cite{Yan-PRL-12,Sun-PRL-2013,Pasqualetti-IEEECNS-2014} and measurement noise\cite{Friedman-Science-04} causes uncertainties in the observation process.
To quantify these issues we formalize the dynamics of a controlled network with $N$ nodes and $N_{\text{D}}$ external control inputs as\cite{Rugh-Book-1996,Sontag-Book-1998,Slotine-Book-91,Liu-Nature-11}
\begin{equation} \label{system2}
\dot{\mathbf{x}}(t) = A\mathbf{x}(t) + B\mathbf{u}(t),
\end{equation}
where the vector $\mathbf{x}(t) = [x_1(t),x_2(t),\ldots, x_N(t)]^\text{T}$ describes the states of the $N$ nodes at time $t$ and $x_i(t)$ can represent the
concentration of a metabolite in a metabolic network\cite{Almaas-Nature-2004},
the geometric state of a chromosome in a chromosomal interaction network\cite{Rajapakse-PNAS-11}, or the belief of an individual in opinion dynamics\cite{Acemoglu-GEB-2010,Castellano-RMP-2009}.
The vector $\mathbf{u}(t) = [u_1(t), u_2(t),\ldots, u_{N_{\text{D}}}(t)]^\text{T}$ represents the external control inputs, and $B$ is the input matrix with $B_{ij} = 1$ if control input $u_j(t)$ is imposed on node $i$.
The adjacency matrix $A$ captures the interactions between the nodes, including the possibility of self-loops $A_{ii}$ representing the self-regulation of node $i$.
\vspace{0.8cm}\\
\noindent{\textbf{Control energy}}\\
The system (1) can be driven from an initial state $\mathbf{x}_{o}$ to any desired final state $\mathbf{x}_{d}$ within the time $t\in[0,\tau]$ using an infinite number of possible control inputs $\mathbf{u}(t)$. The optimal input vector aims to minimize the control energy\cite{Rugh-Book-1996} $\int_0^{\tau}\|\mathbf{u}(t)\|^2dt$, which captures the energy of electronic and mechanic systems or the amount of effort required to control biological and social systems. If at $t$ = 0 the system is in state $\mathbf{x}_o = \mathbf{0}$, the minimum energy required to move the system to point $\mathbf{x}_d$ in the state space can be shown to be\cite{Rugh-Book-1996,Yan-PRL-12,Sun-PRL-2013,Pasqualetti-IEEECNS-2014}
\begin{equation} \label{minienergy}
\mathcal{E}(\tau) = \mathbf{x}^{\text{T}}_{d}G_{c}^{-1}(\tau)\mathbf{x}_{d},
\end{equation}
where $G_{c}(\tau) = \int_0^{\tau}e^{At}BB^{\text{T}}e^{A^{\text{T}}t}dt$ is the symmetric controllability Gramian. When the system is controllable all eigenvalues of $G_{c}(\tau)$ are positive. Eq. (2) indicates that for a network $A$ and an input matrix $B$ the control energy $\mathcal{E}(\tau)$ also depends on the desired state $\mathbf{x}_{d}$. Consequently, driving a network to various directions in the state space requires different amounts of energy. For example, to move the weighted network of Fig. 1a to the three different final states $\mathbf{x}_d$ with $\|\mathbf{x}_d\|=1$, we inject the optimal signals $\mathbf{u}(t)$ shown in Fig. 1b onto node 1, steering the system along the trajectories shown in Fig. 1c. The corresponding minimum energies are shown in Fig. 1d. The control energy surface for all normalized desired states is an ellipsoid, implying that the required energy varies dramatically as we move the system in different directions.

As real systems normally function near a stable state, i.e. all eigenvalues of $A$ are negative\cite{May-Book-1974}, the control energy $\mathcal{E}(\tau)$ decays quickly to a nonzero stationary value when the control time $\tau$ increases\cite{Yan-PRL-12}.
Henceforth we focus on the control energy $\mathcal{E}\equiv\mathcal{E}(\tau \rightarrow \infty )$ and the controllability Gramian $G \equiv G_c(\tau \rightarrow \infty)$.

Given a network $A$ and an input matrix $B$, the controllability Gramian $G$ is unique, embodying all properties related to the control of the system.
To uncover the directions of the state space requiring different energies, we explore the eigen-space of $G$.
Denote by $\mathcal{E}_i$ the eigen-energies, i.e. the minimum energy required to drive the network to $G$'s eigen-directions.
According to Eq. (\ref{minienergy}) $\mathcal{E}_i = 1/\mu_i$ with $\mu_i$ corresponding to $G$'s eigenvalues.
Generally, the energy surface for a network with $N$ nodes is a super-ellipsoid spanned by $G$'s $N$ eigen-energies. To determine the distribution of these eigen-energies we decompose the adjacency matrix as $A = V\Lambda V^{\text{T}}$, where
$V$ represents the eigenvectors of $A$ and $\Lambda = \text{diag}\{-\lambda_1,-\lambda_2,\ldots,-\lambda_N\}$ are the eigenvalues. For stable undirected networks all eigenvalues of $A$ are negative, thus we denote the eigenvalues by $-\lambda_i$ so that the absolute eigenvalues are $\lambda_i > 0$ for all $i$. We sort the absolute eigenvalues in ascending order $0 < \lambda_1 < \lambda_2 < \ldots < \lambda_N$, finding that (SI Sec. I)
\begin{equation}
G = V[(V^{\text{T}}BB^{\text{T}}V)\circ C]V^{\text{T}},
\label{GDecompose}
\end{equation}
where $\circ$ denotes Hadamard product, i.e. $(X\circ Y)_{ij} = X_{ij}Y_{ij}$, and $C$ is a matrix with entries $C_{ij} = \frac{1}{\lambda_i+\lambda_j}$. For a given network, (\ref{GDecompose}) captures the impact of the input matrix $B$ on the control properties of the system, allowing us to analyze the distribution of eigen-energies for different number of driver nodes and determine the required energy for each direction.
\vspace{0.8cm}\\
\noindent\textbf{Controlling a system through all nodes}\\
\noindent If we can control all nodes, i.e. $N_D = N$, $B$ becomes a unit diagonal matrix. In this case $G = V\text{diag}\{\frac{1}{2\lambda_i}\}V^{\text{T}}$ and the eigen-directions of the controlled system are the same as the network's eigenvectors. Thus $\mathcal{E}_{i} = 2\lambda_i$ and $p(\mathcal{E}) = (1/2)p(\lambda)$, i.e. the distribution of eigen-energies is proportional to the distribution of the network's absolute eigenvalues. We add self-loops as $A_{ii} = -(\delta + \sum_{j=1}^{N}A_{ij})$ where $\delta > 0$ is a small perturbation to ensure that all eigenvalues of $A$ are negative. This scheme has been widely used in previous studies on dynamical processes taking place on networks, such as opinion dynamics\cite{Acemoglu-GEB-2010}, synchronization\cite{Pecora-PRL-1998}, and control\cite{Yan-PRL-12}. For networks with degree distribution\cite{Albert-RMP-02,Cohen-Book-10,Newman-Book-10} $p(k) \sim k^{-\gamma}$ the distribution of $A$'s absolute eigenvalues also obeys a power law\cite{Chung-PNAS-2003,Kim-Chaos-2007} $p(\lambda) \sim \lambda^{-\gamma}$ (see SI Sec. II A). Consequently,
\begin{equation}
p(\mathcal{E}) \sim \mathcal{E}^{-\gamma},
\label{Nd=N}
\end{equation}
indicating that the system can be easily driven in most directions of the state space, requiring a small $\mathcal{E}$. A few directions require considerable energy and the
most difficult direction needs\cite{Cohen-PRL-00} $\mathcal{E}_{\text{max}} \sim N^{\frac{1}{\gamma-1}}$.
The fact that $\mathcal{E}_{\text{max}}$ is sub-linear in $N$ for $\gamma > 2$ indicates that, when $N_{\text{D}}=N$, the energy density $\mathcal{E}/N_{\text{D}}$ remains bounded. In Fig. 2 we test the prediction (\ref{Nd=N}) for several model, infrastructural, social, and biological networks. We find that $p(\mathcal{E})$ follows a power law for uncorrelated or correlated scale-free model networks (Figs. 2a-b) , the airline transportation network (Fig. 2d), the Internet AS-level network (Fig. 2e), an Israeli social network (Fig. 2g), the user-interaction network of an online forum (Fig. 2h), the human protein-protein interaction network (Fig. 2j), and the human heterogeneous network (Fig. 2l) in line with the prediction (\ref{Nd=N}). In contrast, for several networks with bounded degree distribution, like the Erd\H{o}s-R\'enyi random network (Fig. 2c), the US power grid network (Fig. 2f), the interlocking network of Norwegian companies (Fig. 2i), and the functional coactivation network of the human brain (Fig. 2l), $p(\mathcal{E})$ is also bounded, as predicted by $\gamma \rightarrow \infty$ in (\ref{Nd=N}). Such networks require even less energy for controlling their progress in their most difficult direction. Taken together, we find that for $N_{\text{D}} = N$ the distribution of eigen-energies is uniquely determined by network topology and we lack significant energetic barriers for control.
\vspace{0.8cm}\\
\noindent {\textbf{Controlling a system through a single node}}\\
\noindent If all nodes exhibit nonidentical self-loops we can control an undirected network by driving only a single node\cite{Cowan-PL-12,Yuan-NC-2013}. In this case $V^{\text{T}}BB^{\text{T}}V = \{V_{ih}V_{jh}\} \sim \mathcal{O}(1/N)$, where $h$ is the index of the chosen driver node. Thus $V^{\text{T}}BB^{\text{T}}V$ can be viewed as a small perturbation to the matrix
$C$ in (\ref{GDecompose}). The statistical behavior of $G$'s eigenvalues is mainly determined by the eigenvalues of $C$, which can be approximated by Cholesky factors\cite{Antoulas-Book-2009}.
As mentioned above, for networks with $p(k) \sim k^{-\gamma}$, the distribution of $A$'s absolute eigenvalues also follows a power law, providing the $i$-th eigenvalue $\lambda_i \sim (\frac{N}{N+1-i})^{\frac{1}{\gamma-1}}$ (SI Sec. IIB). If $\gamma \rightarrow 0$, i.e.
for extremely heterogeneous networks\cite{Charo-PRL-2011,Nepusz-NP-12}, the eigenvalue gaps $g_{i}\equiv \lambda_{i+1}-\lambda_i$ are
identical. For $\gamma \rightarrow \infty$ (homogeneous networks), $g_i$ is again uniform. Hence it is reasonable to assume $g_i = g$ for all $i$, allowing us to analytically obtain the distribution of eigen-energies as $p(\mathcal{E}) \sim 1/(1 + 1/\mathcal{E})\mathcal{E}^{-1}$ (see SI Sec. III A, B). Therefore,
\begin{equation}
p(\mathcal{E}) \sim \mathcal{E}^{-1}
\label{Nd=1}
\end{equation}
for large $\mathcal{E}$.
Equation (5) predicts that, to drive a stable network of $N$ nodes with a single driver node, the most difficult direction in the state space requires
$\mathcal{E}_{\text{max}} \sim e^{N}$ energy (SI Sec. III C). This exponential $N$-dependence makes the control of large networks in the most difficult direction energetically infeasible. For validation we also consider the complementary
cumulative distribution $p_>(\mathcal{E}) = \int_{\mathcal{E}}^{\mathcal{E}_{\text{max}}}p(\mathcal{E}')d\mathcal{E}'$. Based on (\ref{Nd=1}) we obtain $p_>(\mathcal{E}) \sim (\ln{\mathcal{E}_{\text{max}}}-\ln{\mathcal{E}})$, decreasing linearly with $\ln{\mathcal{E}}$.
We test our prediction on several network models (Figs. 3a-c) and real networks (Figs. 3d-l), finding that the corresponding eigen-energies span over a hundred orders of magnitude and this exceptional range of variations are reasonably well approximated by (\ref{Nd=1}) for both $p_>(\mathcal{E})$ and $p(\mathcal{E})$. Taken together, if we attempt to control a network from a single node ($N_{\text{D}} = 1$), the required energy varies enormously for different directions, almost independently of the network structure, making some directions prohibitively expensive energetically.
\vspace{0.8cm}\\
\noindent {\textbf{Controlling a system through a finite fraction of its nodes}}\\
\noindent When $p(\mathcal{E}) \sim \mathcal{E}^{-\gamma}$, the distribution $p(\mathcal{\hat{E}}) \sim e^{(1-\gamma)\mathcal{\hat{E}}}$ where $\mathcal{\hat{E}}\equiv \ln\mathcal{E}$. Thus, if $N_{\text{D}} = N$, $p(\mathcal{\hat{E}})$ is an exponential (one-peak) distribution for $\gamma > 2$ in (\ref{Nd=N}); if $N_{\text{D}} = 1$, as $p(\mathcal{E}) \sim \mathcal{E}^{-1}$ in (\ref{Nd=1}), $p(\mathcal{\hat{E}})$ is a uniform distribution.
To understand the transition from (\ref{Nd=N}) for $N_{\text{D}}=N$ to (\ref{Nd=1}) for $N_{\text{D}}=1$, we investigate the
distribution $p(\mathcal{\hat{E}})$ when $1<N_{\text{D}}<N$, i.e. when we try to control a system through a finite fraction of its nodes. In this case we find that $p(\hat{\mathcal{E}})$ has multiple peaks (Fig. 4a), which are induced by gaps in the eigen-energy spectrum (Fig. 4b).
For $N_{\text{D}}/N=0.6$, a gap separates the eigen-energies into two bands, such that the lower band contains $N_{\text{D}}$ eigen-energies.
This gap leads to two peaks in the distribution $p(\hat{\mathcal{E}})$ as shown in Fig. 4c. When we have fewer driver nodes ($N_{\text{D}}/N$ decreases),
the number of peaks $N_{\text{peak}}$ increases (Fig. 4a). We find that $N_{\text{peak}} = \text{int}[N/N_{\text{D}}]$, predicting $N_{\text{peak}} = 2, 4, 5$ for $N_{\text{D}}/N = 0.5, 0.25, 0.2$, respectively (see also SI Fig. S3).
The multi-peak nature of $p(\mathcal{\hat{E}})$ has two important implications. First, the boundary of the first energy
band $\mathcal{E}_{N_{\text{D}}}$ varies only weakly with $N_{\text{D}}$ (Fig. 4d), indicating that the energy required
to move the network within the subspace spanned by the first $N_{\text{D}}$ eigen-directions is relatively small. Second, $\mathcal{\hat{E}}$ (i.e. $\log\mathcal{E}$) grows linearly from one band to the next (SI Fig. S4). Thus,
$\log \mathcal{E}_{\text{max}}$ (the boundary of the last band) is linearly dependent on the number of peaks, i.e. $\mathcal{E}_{\text{max}} \sim e^{N/N_{\text{D}}}$ (Fig. 4d). Controlling a single node induces $N$ peaks in $p(\mathcal{\hat{E}})$, consequently the distribution $p(\mathcal{\hat{E}})$ becomes uniform (SI Fig. S3), resulting in $p(\mathcal{E}) \sim \mathcal{E}^{-1}$ of (\ref{Nd=1}) and $\mathcal{E}_{\text{max}} \sim e^{N}$. We numerically test the prediction $\mathcal{E}_{\text{max}} \sim e^{N/N_{\text{D}}}$ for several real networks (Figs. 4e-j), the result being in excellent agreement with our prediction.

In Table 1 we summarize our findings about the distribution of eigen-energies and the maximum energy required to control a system towards the most difficult direction.
\vspace{0.8cm}\\
\noindent\textbf{Implications to observational uncertainty}\\
\noindent
The results obtained above have direct implications for observability as well. Indeed,
consider a system governed by the dynamics
\begin{align}
    \dot{\mathbf{x}}(t) &= A\mathbf{x}(t) \\
    \mathbf{y}(t) &= C\mathbf{x}(t) + \mathbf{w}(t)
\end{align}
with an initial state $\mathbf{x}_o \neq 0$, where $C$ is the output matrix and $\mathbf{y}(t)$ are the output signals including measurement noise $\mathbf{w}(t)$, which we assume to be a Gaussian white noise with zero mean and variance one. We aim to estimate $\mathbf{\hat{x}}_o$ of the initial state $\mathbf{x}_o$ while minimizing the difference $\int_0^{\tau}\|\mathbf{y}(t)-\mathbf{\hat{y}}(t)\|^2dt$ between the output $\mathbf{y}(t)$ that is actually observed and the output $\hat{\mathbf{y}}(t)=Ce^{At}\hat{\mathbf{x}}_o$ that would be observed in the absence of noise.
With the maximum-likelihood approximation\cite{Kailath-Linear-2000}, the expectation $\langle \hat{\mathbf{x}}_o \rangle = \mathbf{x}_o$ and
the covariance matrix\cite{Kailath-Linear-2000} $\langle \tilde{\mathbf{x}}\tilde{\mathbf{x}}^{\text{T}} \rangle= G_{o}^{-1}(\tau)$,
where $\tilde{\mathbf{x}} \equiv \hat{\mathbf{x}}_o - \mathbf{x}_o$ is estimation error and
$G_{o}(\tau) = \int_0^{\tau}e^{A^{\text{T}}t}C^{\text{T}}Ce^{At}dt$ is the observability Gramian.
Therefore, the variance $\sigma^2$ of the approximation in direction $\tilde{\mathbf{x}}$ is
\begin{equation}
\sigma^2(\tau) = \tilde{\mathbf{x}}^{\text{T}}G_{o}^{-1}(\tau)\tilde{\mathbf{x}},
\label{Go}
\end{equation}
indicating that the estimation uncertainty varies with the direction of the state space.
To illustrate this, consider the network in Fig. 1e that moves along the trajectory of Fig. 1g, while we measure the state of the sensor node and plot the noisy output $\mathbf{y}(t)$ in Fig. 1f. With the maximum-likelihood approximation we reconstruct $\mathbf{x}_o$ from $\mathbf{y}(t)$ and show the estimation error $\tilde{\mathbf{x}} \equiv \hat{\mathbf{x}}_o - \mathbf{x}_o$ for thousands of independent runs (Fig. 1h). The estimation variance is different for various directions, forming an uncertainty ellipsoid. Thanks to the duality between $G_c(\tau)$ and $G_o(\tau)$, the control energy for a direction in Fig. 1d represents the estimation variance for the same direction in Fig. 1h.

To be specific, due to the duality of the controllability Gramian $G_{c}$ in (\ref{minienergy}) and the observability Gramian $G_o$ in (\ref{Go}), we have $\sigma^2 = \mathcal{E}$ for the same direction, implying that the least controllable direction (i.e. the direction requiring the most energy) is also least observable (having highest uncertainty). Therefore, our findings about the distribution of eigen-energies apply directly to the distribution of $\sigma^2$ along the eigen-directions: if all nodes are sensor nodes ($N_{\text{S}} = N$) we have $p(\sigma^2) \sim (\sigma^2)^{-\gamma}$; if we attempt to observe the system from a single node ($N_{\text{S}} = 1$) we have $p(\sigma^2) \sim (\sigma^2)^{-1}$; and for a finite fraction of sensor nodes ($1 < N_{\text{S}} < N$) the largest observational uncertainty $\sigma^2_{\text{max}}$ decreases exponentially when the number of sensor nodes increases, i.e. $\sigma^2_{\text{max}} \sim e^{N/N_{\text{S}}}$.
\vspace{0.8cm}\\
\noindent\textbf{Beyond the degree distribution}\\
Real networks have a number of additional properties that are not encoded by their degree distributions, like local clustering\cite{Watts-Nature-98},
degree correlations\cite{Newman-PRL-02}, and community structure\cite{Girvan-PNAS-2002} (SI Table S1). To assess the impact of these topological
characteristics we perform the degree-preserved randomization\cite{Xulvi-Brunet-PRE-2004} on each network, eliminating local clustering, degree
correlations and modularity. We find that the distribution of eigen-energies required to drive each randomized network follows the predictions (4)
and (5) (Figs. S6 and S7 in SI), indicating that degree distribution is the main factor determining $p(\mathcal{E})$. When the number of driver nodes increases, the maximal control energy for the randomized networks decreases exponentially, as predicted earlier (Fig. S8 in SI). We also validate the predictions (4) and (5) on model networks with positive or negative degree correlations (Figs. 1b and 2b). All the tests indicate that the strength of local clustering, degree correlations or community structure have only minor influence on the behavior of control energy. Consequently, our calculations for uncorrelated networks capture the correct fundamental dependence of control energy for real networks.

Many real networks have dead ends, i.e. nodes with one degree, which can undermine the stability of complex systems\cite{Menck-NC-2014}. To test the impact of dead ends on control energy we explored several real networks that contain a considerable number of one-degree nodes (see SI Table S1). As shown in Figs. 2-4 and S6-S8 the predictions are robust against such dead ends.
\vspace{0.8cm}\\
\noindent\textbf{Conclusion and discussion}\\
\noindent The energy required for control is a significant issue for practical control of complex systems. By exploring the eigen-space of controlled systems we found that if all nodes of a system are directly driven, the eigen-energies can be heterogeneous or homogeneous, depending on the structure of underlying networks. Yet, if we wish to control a system through a single node, the eigen-energies are enormously heterogeneous, almost independently of the network structure. Finally, if a finite fraction of nodes are driven, the maximum control energy decays exponentially with the increasing number of driver nodes. Taken together, our results indicate that even if controllable, most systems still have directions which are energetically inaccessible, suggesting a natural mechanism to avoid undesirable states. Indeed, many complex systems, such as transcriptional networks for gene expression\cite{Muller-NatureComment-11} and sensorimotor systems for motion control\cite{Todorov-NatureNeuRev-02}, only need to function in a low-dimensional subspace. Due to the duality of controllability and observability, our results also imply that, if we monitor only a small fraction of nodes, the observation can be extremely unreliable in certain directions of the phase space.

It is worth noting that linear dynamics captures the behavior of nonlinear systems in the vicinity of their equilibria. The formalism (1) has been widely used to model diverse networked systems\cite{Rajapakse-PNAS-11,Acemoglu-GEB-2010,Castellano-RMP-2009,Pasqualetti-IEEECNS-2014,Tzoumas-arXiv-2015} (see also SI Sec. VII A, B), allowing us to reveal the role of the network topology on the fundamental control properties of complex systems\cite{Liu-Nature-11,Rajapakse-PNAS-11,Nepusz-NP-12,Yan-PRL-12,Sun-PRL-2013,Pasqualetti-IEEECNS-2014,Tang-PO-2012,Jia-NC-2013,Yuan-NC-2013,Ruths-Science-2014,Giulia-PRL-2014,summers2014submodularity,Tzoumas-arXiv-2015}. Indeed, if the linearized system (1) is controllable, the original nonlinear system is locally controllable\cite{Coron-Book-2009}. The corresponding control energy is also highly heterogeneous for different directions, if we constrain the system's trajectory to be local (SI Sec. VII C). Moreover, if the linearized dynamics of a nonlinear system is controllable along a specific trajectory, the original nonlinear system is locally controllable along the same trajectory\cite{Coron-Book-2009}. This implies that our results can be potentially extended to describe control properties of nonlinear systems in the vicinity of their stability basin\cite{Menck-NP-2013,Menck-NC-2014}. Yet, in this case, the linearized dynamics becomes time-varying, and the required energy for controlling time-varying systems remains an open problem that deserves future attention.

\vspace{0.6cm}
\noindent\textbf{Acknowledgements}\\
\noindent We thank E. Guney, C. Song, J. Gao, M. T. Angulo, S. P. Cornelius, B.  Coutinho, and A. Li for discussions. This work was supported by Army Research Laboratories (ARL) Network Science (NS)
Collaborative Technology Alliance (CTA) grant: ARL NS-CTA W911NF-09-2-0053; DARPA Social Media in Strategic Communications project under agreement number W911NF-12-C-002; the John Templeton Foundation: Mathematical and Physical Sciences grant no. PFI-777; European Commission grants no. FP7 317532 (MULTIPLEX) and 641191 (CIMPLEX).
\vspace{0.6cm}\\
\noindent\textbf{Author contributions}\\
All authors designed and performed the research. G.Y. and G.T. carried out the numerical calculations. G.Y. did the analytical calculations and analysed the empirical data. G.T., B.B., J.-J.S., Y.-Y.L. and A.-L.B. analysed the results. G.Y. and A.-L. B. were the main writers of the manuscript. G.T., B.B. and Y.-Y.L. edited the manuscript. G.Y. and G.T. contributed equally to this work.
\vspace{0.6cm}\\
\noindent\textbf{Corresponding Author}\\
Correspondence and requests for materials should be addressed to A-L.B. (email: alb@neu.edu).

\begin{figure}
\centering
\includegraphics[width=0.67\textwidth]{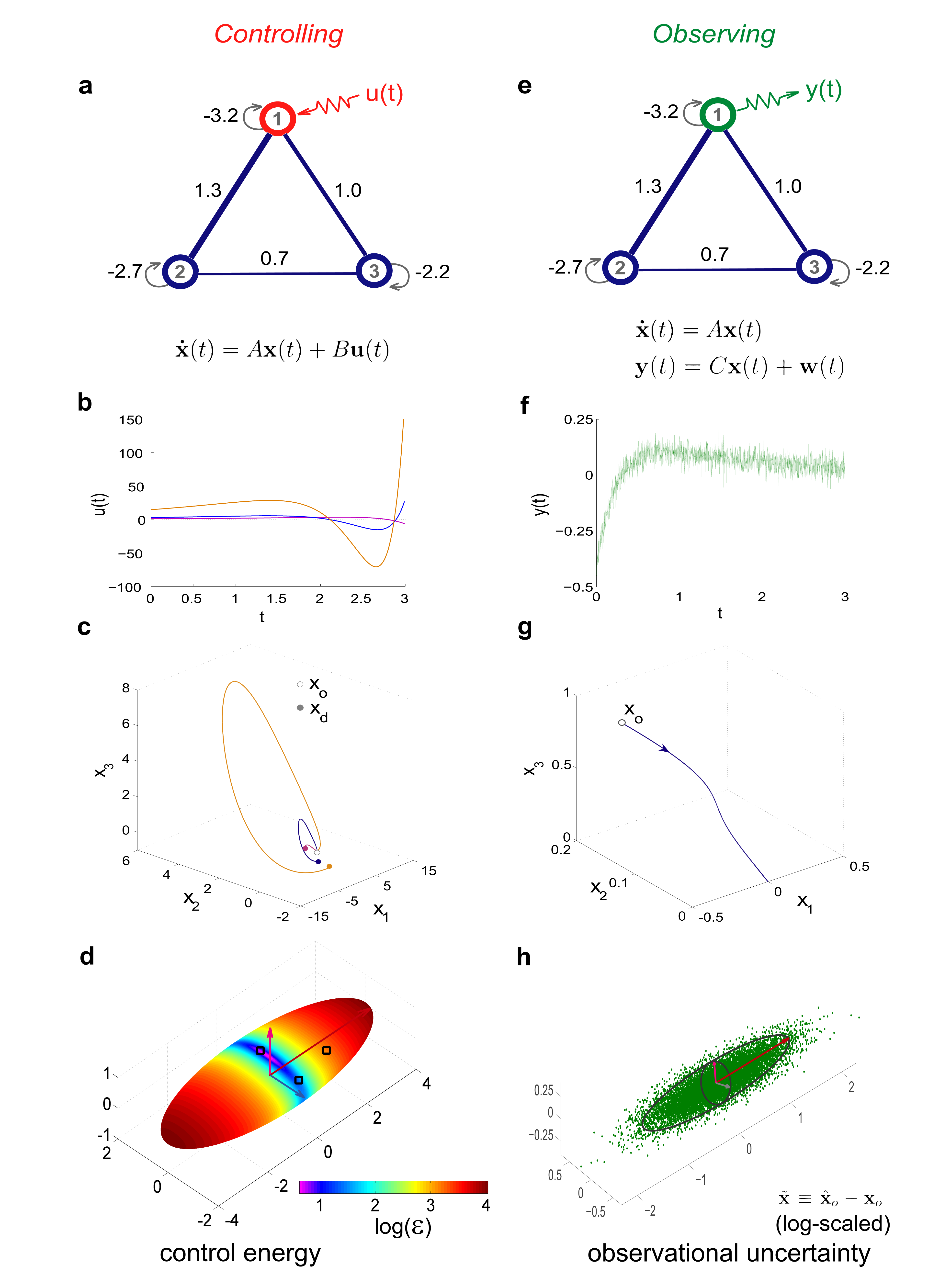}
\caption{Controlling and observing a network. \textbf{a}, The control of a three-node weighted network with one external signal $u(t)$ that is injected to the red driver node. Hence the input matrix is $B=[1,0,0]^{\text{T}}$. The nodes have negative self-loops which make all eigenvalues of the adjacency matrix $A$ negative. \textbf{b}, Optimal control signals which minimize the energies required to move the network from the initial state $\mathbf{x}_{o}=[0,0,0]^{\text{T}}$ to three different desired states $\mathbf{x}_d$ with $\|\mathbf{x}_d\| = 1$ in the given time interval $t \in [0,3]$. \textbf{c}, The trajectories of the network state $\mathbf{x}(t)$ driven respectively by the control signals in \textbf{b}. \textbf{d}, The control energy surface, showing the amount of energy required to move the network by one unit distance (i.e. $\|\mathbf{x}_d\| = 1$) in different directions. The surface is an ellipsoid spanned by the eigen-energies for the controllability Gramian's three eigen-directions (arrows). The squares correspond to the three final points used in \textbf{b} and \textbf{c}. \textbf{e}, Observing the network with one output $y(t)$. Node 1 is selected as the sensor (green), thus the output matrix is $C=[1,0,0]$. The measurement noise $\mathbf{w}(t)$ is assumed as Gaussian white noise with zero mean and variance one. \textbf{f}, A typical output $y(t)$ that is used to approximate the initial state $\mathbf{x}_o$. \textbf{g}, A typical trajectory of the system state $\mathbf{x}(t)$. \textbf{h}, Estimation error $\mathbf{\tilde{x}} = \hat{\mathbf{x}}_o - \mathbf{x}_o$, where $\hat{\mathbf{x}}_o$ is the maximum-likelihood estimator of the initial state. Starting from the same initial state we ran the system $5,000$ times independently, each dot representing the estimation error of one run. The uncertainty ellipsoid (black) corresponds to the standard deviation of $\mathbf{\tilde{x}}$ in any direction.}
\label{fig1}
\end{figure}

\begin{figure}
\centering
\includegraphics[width=0.80\textwidth]{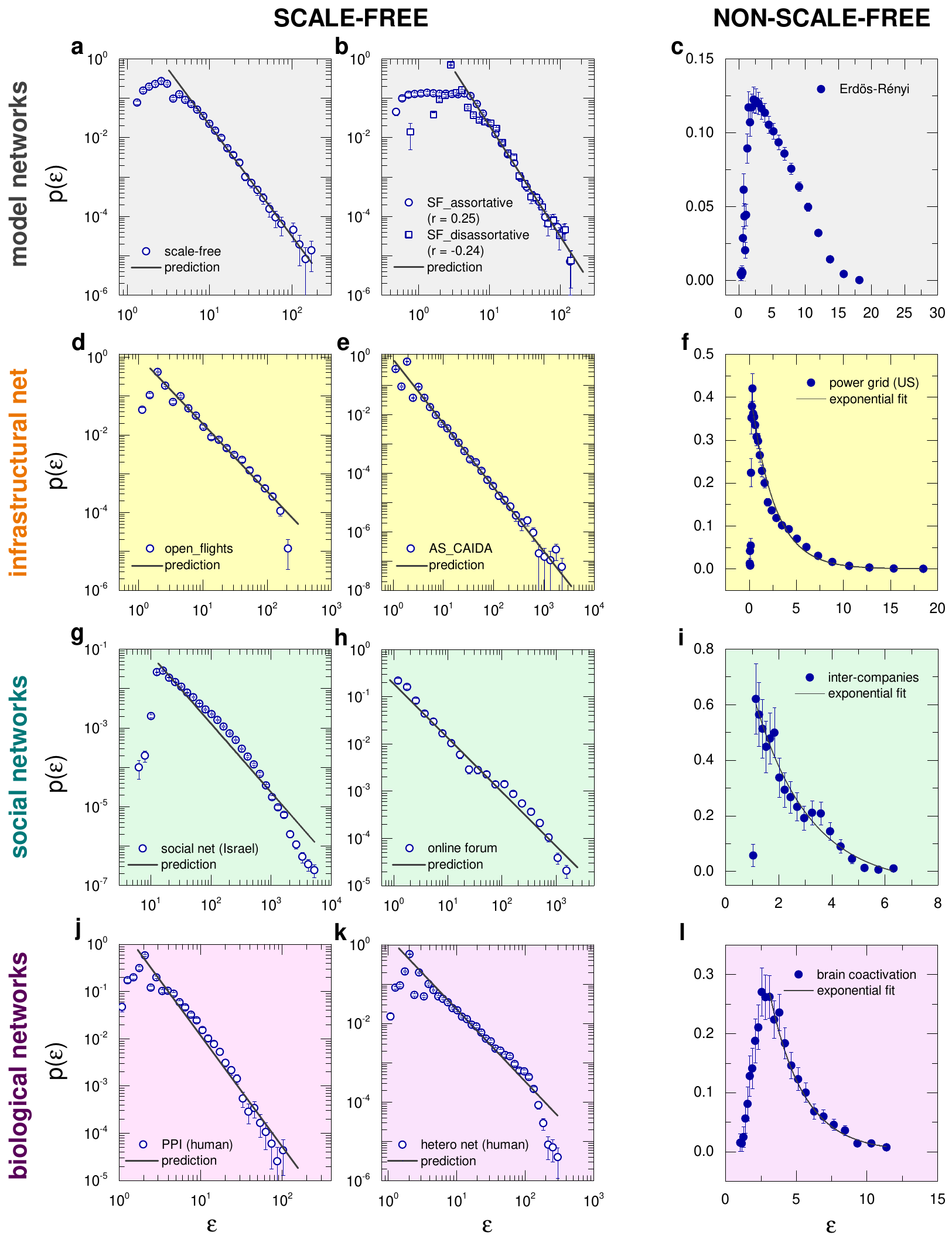}
\caption{Controlling a network through all nodes ($N_\text{D}=N$). The panels show the distribution $p(\mathcal{E})$ of eigen-energies required to control several model and real systems: \textbf{a}, a scale-free model network without degree-degree correlation; \textbf{b}, scale-free model networks with positive ($r=0.25$) or negative ($r=-0.24$) degree-degree correlation; \textbf{c}, an Erd\H{o}s-R\'enyi model network;
\textbf{d}, the airline transportation network; \textbf{e}, the Internet AS-level network; \textbf{f}, the US power grid network; \textbf{g}, an Israeli social network; \textbf{h} the user-interaction network of an online forum; \textbf{i}, the interlocking network of Norwegian companies; \textbf{j}, the human protein-protein interaction network; \textbf{k}, the human heterogeneous network; and \textbf{l}, the functional coactivation network of the human brain. The straight lines show prediction (\ref{Nd=N}) and the error bars represent standard deviations.
For model networks the edges' weights $A_{ij}$ are uniformly drawn from $[0,1]$. The self-loops $A_{ii} = - \sum_j{A_{ij}}-\delta$ where $\delta=0.25$, representing a small perturbation to diagonal entries, to ensure that the network is stable. The data sources and basic characteristics of these networks are discussed in SI Sec. VI A.}
\label{fig2}
\end{figure}

\begin{figure}
\centering
\includegraphics[width=0.77\textwidth]{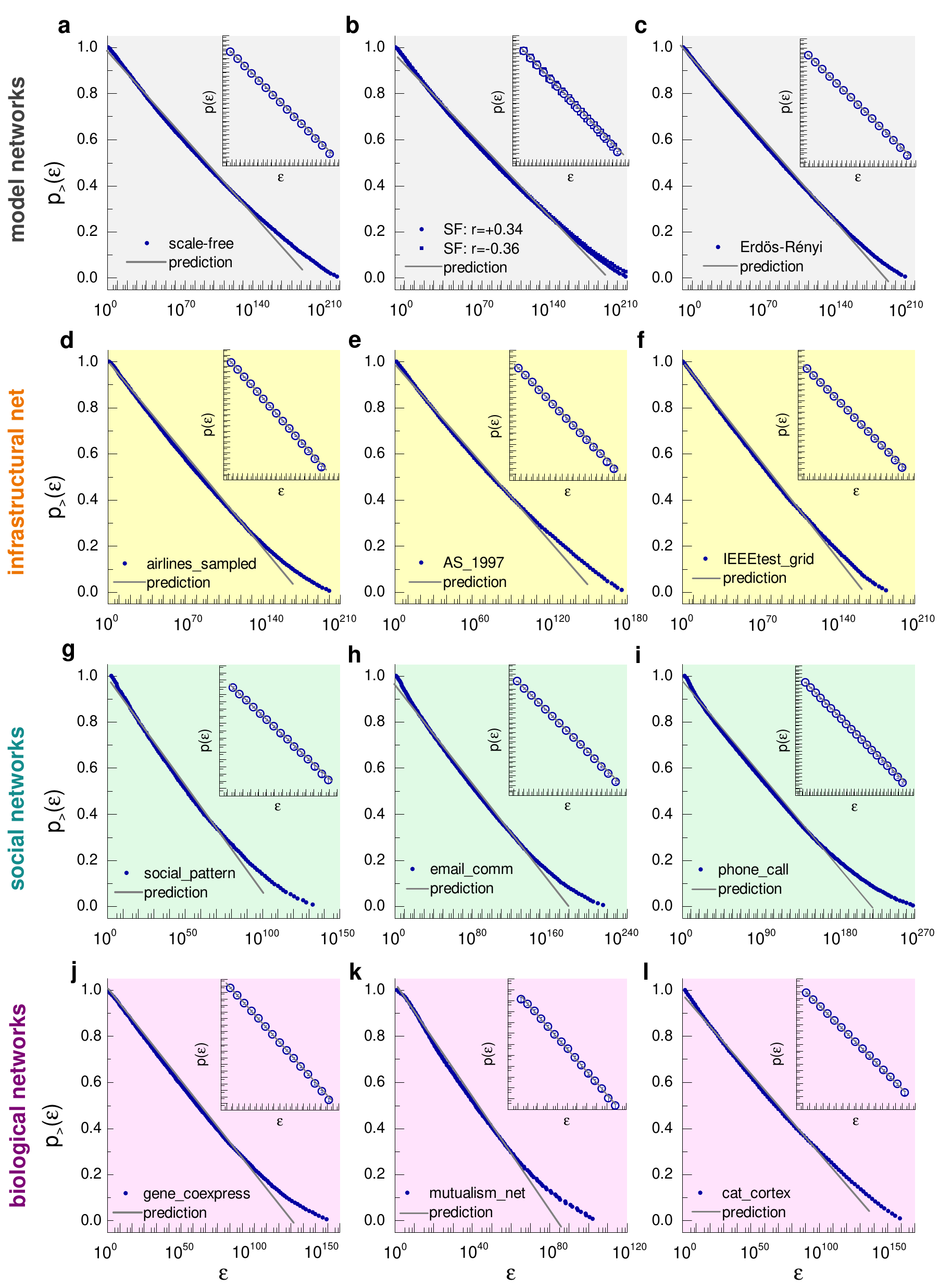}
\caption{
Controlling a network through a single node ($N_\text{D}=1$).
The panels show the complementary cumulative distribution $p_>(\mathcal{E})$ of eigen-energies required to control several model and real systems: \textbf{a}, a scale-free model network without degree-degree correlation; \textbf{b}, scale-free model networks with positive ($r=0.34$) or negative ($r=-0.36$) degree-degree correlation; \textbf{c}, an Erd\H{o}s-R\'enyi model network; \textbf{d}, a sampled airline network; \textbf{e}, an Internet AS-level network in 1997; \textbf{f}, an IEEE power grid test network; \textbf{g}, the human-contact network of the ACM Hypertext 2009 conference; \textbf{h}, an email interaction network; \textbf{i}, the phone call network between different countries; \textbf{j}, a connected component of the human gene-coexpression network; \textbf{k}, a mutualism ecological network in Mauritius; and \textbf{l}, the inter-region network of cat cortex. The insets are the log-log plots of probability distributions $p(\mathcal{E})$ with logarithmic binning\cite{Newman-Book-10}. The straight lines show prediction (\ref{Nd=1}) and the error bars represent standard deviations. The data sources and basic characteristics of these networks are discussed in SI Sec. VI A.}
\label{fig3}
\end{figure}

\begin{figure}
\centering
\includegraphics[height=\textwidth]{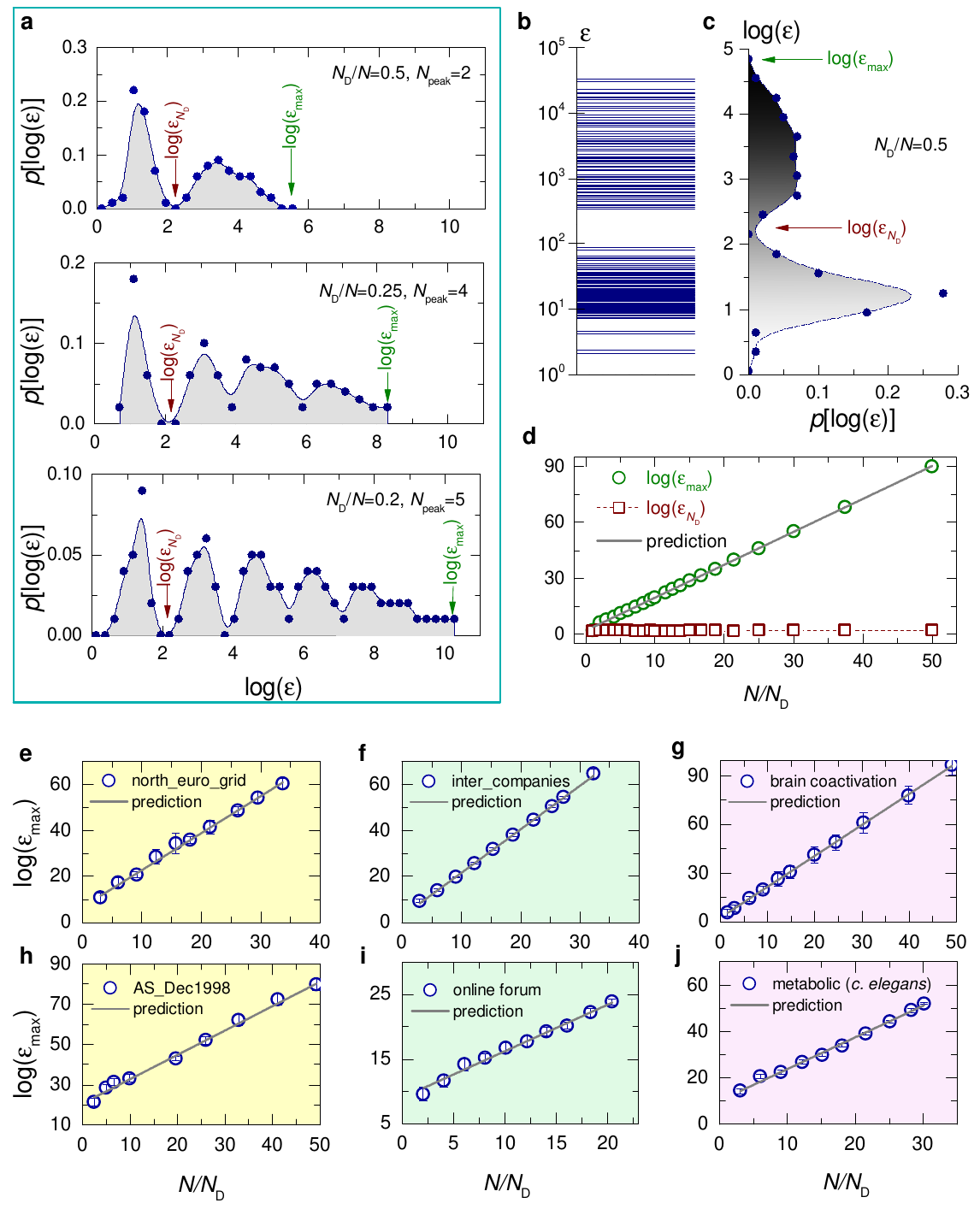}
\caption{
Controlling a network through a finite fraction of its nodes.
\textbf{a}, The multi-peak distributions $p[\log(\mathcal{E})]$ for $1<N_{\text{D}}<N$, where the dots represent numerical results. The solid curves and shaded areas are smoothed for illustration. $\mathcal{E}_{N_{\text{D}}}$ denotes the boundary of the first energy band that contains $N_{\text{D}}$ eigen-energies. $\mathcal{E}_{\text{max}}$ is the maximum control energy corresponding to the most difficult direction. The number of peaks is $N_{\text{peak}}$ $= \text{int}[N/N_{\text{D}}]$ (see also SI Fig. S3). \textbf{b}, The eigen-energy spectrum for controlling the network. There is a gap in the logarithmic scale between the $N_{\text{D}}$-th and the $(N_{\text{D}}+1)$-th smallest eigen-energies, which leads to the two-peak distribution $p[\log(\mathcal{E})]$ in \textbf{c}. \textbf{d}, $\log(\mathcal{E}_{\text{max}})$ and $\log(\mathcal{E}_{N_{\text{D}}})$ as functions of $N/N_{\text{D}}$, indicating that $\mathcal{E}_{\text{max}} \sim e^{N/N_{\text{D}}}$ while $\mathcal{E}_{N_{\text{D}}}$ depends weakly on $N_{\text{D}}$. We also test the prediction in real networks: \textbf{e}, the North European power grid network; \textbf{f}, the interlocking network of Norwegian companies; \textbf{g}, the functional coactivation network of the human brain; \textbf{h}, an Internet AS-level network in Dec 1998; \textbf{i}, the user-interaction network of an online forum; and \textbf{j}, the metabolic network of {\it C. elegans}. The straight lines show the prediction $\mathcal{E}_{\text{max}} \sim e^{N/N_{\text{D}}}$ and the error bars represent standard deviations. The data sources and basic characteristics of these networks are discussed in SI Sec. VI A.}
\label{fig4}
\end{figure}

\newpage
\begin{center}
\textbf{Table 1: Controlling complex networks with different number of driver nodes}\\
\begin{tabular}{ |l| l| l| }
  \specialrule{0.8pt}{1pt}{1pt}
  \textbf{Number of driver nodes}& \textbf{Distribution of eigen-energies}& \textbf{Maximum control energy}\\
  \specialrule{0.8pt}{1pt}{1pt}
  $N_{\text{D}}=N$ & $p(\mathcal{E}) \sim \mathcal{E}^{-\gamma}$ &  $\mathcal{E}_{\text{max}} \sim N^{\frac{1}{\gamma-1}}$ \\ \hline
  $1 < N_{\text{D}} < N$ & $N_{\text{peak}} = \text{int}[N/N_{\text{D}}]$ for $p[\log(\mathcal{E})]$& $\mathcal{E}_{\text{max}} \sim e^{N/N_{\text{D}}}$ \\ \hline
  $N_{\text{D}} = 1$ & $p(\mathcal{E}) \sim \mathcal{E}^{-1}$ & $\mathcal{E}_{\text{max}} \sim e^{N}$\\
  \specialrule{0.8pt}{1pt}{1pt}
\end{tabular}
\end{center}
$N$ is the total number of nodes and $N_{\text{D}}$ is the number of driver nodes. $\gamma$ is the exponent of the degree distributions $p(k) \sim k^{-\gamma}$. For large $\gamma$ the network becomes degree-homogeneous, behaving similarly to a random network.

\end{document}